\documentclass[fleqn,10pt]{wlscirep}
\usepackage[utf8]{inputenc}
\usepackage[T1]{fontenc}
\title{Femtosecond photoelectron circular dichroism of chemical reactions}
\usepackage{setspace} 
\usepackage[version=3]{mhchem}
\usepackage{xspace}
\usepackage{upgreek} 

\author[1,*]{V\'it Svoboda}
\author[1,3]{Niraghatam Bhargava Ram}
\author[1]{Denitsa Baykusheva}
\author[1]{Daniel Zindel}
\author[1]{Max D.J. Waters}
\author[2]{Benjamin Spenger}
\author[1]{Manuel Ochsner}
\author[1]{Holger Herburger}
\author[2]{J\"urgen Stohner}
\author[1,*]{Hans Jakob W\"orner}
\affil[1]{Laboratory of Physical Chemistry, ETH Z\"urich, 8093 Z\"urich, Switzerland}
\affil[2]{Institute of Chemistry and Biotechnology, Zurich University of Applied Sciences, 8820 W\"adenswil, Switzerland}
\affil[3]{Department of Physics, Indian Institute of Science Education and Research - Bhopal, Bhauri, Bhopal 462066, India}

\affil[*]{vit.svoboda@phys.chem.ethz.ch, hwoerner@ethz.ch}

\begin{abstract}
Understanding the chirality of molecular reaction pathways is essential for a broad range of fundamental and applied sciences. However, the current ability to probe chirality on the time scale of chemical reactions remains very limited. Here, we demonstrate time-resolved photoelectron circular dichroism (TRPECD) with ultrashort circularly polarized vacuum-ultraviolet (VUV) pulses from a table-top source. We demonstrate the capabilities of VUV-TRPECD by resolving the chirality changes in time during the photodissociation of atomic iodine from two chiral molecules. We identify several general key features of TRPECD, which include the ability to probe dynamical chirality along the complete photochemical reaction path, the sensitivity to the local chirality of the evolving scattering potential, and the influence of electron scattering off dissociating photofragments. Our results are interpreted by comparison with novel high-level {\it ab-initio} calculations of transient PECDs from molecular photoionization calculations. Our experimental and theoretical techniques define a general approach to femtochirality.
\end{abstract}
\begin{document}

\flushbottom
\maketitle
%
%
\thispagestyle{empty}


\section*{Introduction}

Most molecules of biochemical relevance are chiral. The mechanisms controlling the chiral selectivity of chemical reactions have therefore been intensively studied over many decades. In spite of fundamental advances, our ability to track the dynamically evolving chirality of reacting molecules remains surprisingly limited. The most mature techniques of chiral recognition rely on tiny differences in the absorption of circularly polarized light by the enantiomers of a chiral molecule, an effect known as circular dichroism (CD), which arises from the interference of electric and magnetic dipole transitions. Considerable efforts have been invested in the development of time-resolved CD (TRCD) methods, which have been successfully applied to liquid systems, as reviewed in Ref.~\cite{Meyer-Ilse2012} and recently in Ref.~\cite{Oppermann2019}. The latter overcomes some initial struggles to achieve the desirable sensitivity and temporal resolution because of the inherent weakness of the CD effect. 

This situation has motivated a recent surge of research activity to find more sensitive methods for chiral discrimination. These recent developments include microwave three-wave mixing \cite{Patterson2013, Hirota2012}, Coulomb-explosion imaging \cite{Pitzer2013, Herwig2013}, high-harmonic generation (HHG) in weakly elliptical \cite{Cireasa2015} and bicircular laser pulses \cite{Harada2018,Baykusheva2018}, and photoelectron circular dichroism (PECD) \cite{Ritchie1976, Powis2000, Bowering2001}. Among these, PECD has been applied to probe the photophysical relaxation dynamics in highly electronically excited (Rydberg) states \cite{Comby2016, Beauvarlet2022} and HHG in bicircular fields has been used to probe a photochemical reaction \cite{Baykusheva2019}. Most recently, the Coulomb explosion of a core-ionized chiral molecule has been observed at LCLS (Linac Coherent Light Source) \cite{Ilchen2021}, but the PECD measured at three pump-probe delays did not display significant variations, in agreement with theory. 

Here, we report the realization of TRPECD with vacuum-ultraviolet (VUV) pulses, demonstrating a general probe of the chirality of photochemical dynamics. This technique combines the benefits of VUV time-resolved photoelectron spectroscopy \cite{Nugentglandorf2001, Squibb2018, Conta2018, Svoboda2019} with the chiral sensitivity of PECD. The photon energy is sufficiently high for ionization along the entire reaction pathway from the excited state through all intermediate configurations to the final photoproducts. Additionally, depending on the ionization energy of the unexcited molecules, the photon energy of the probe may be sufficiently low to prevent their ionization, yielding a background-free technique. Our technique transposes the remarkable chiral sensitivity of PECD from probing static molecules in their electronic ground or excited states  \cite{Powis2008, Powis2014, Powis2014a, Powis2008a, Garcia2014, Garcia2013, Hergenhahn2004, Lux2016, Lux2015, Lux2012, Miles2017, Kastner2016, Kastner2017,Goetz2017, Artemyev2015} to probing chiral light-induced dynamics in electronically excited states. This extension opens many promising avenues since PECD effects have been reported for valence orbitals all the way down to core orbitals \cite{Powis2000, Powis2008, Hergenhahn2004, Ulrich2008, Lehmann2013}, they have been shown to be sensitive to molecular conformation and chemical substitution \cite{Garcia2014, Nahon2016, Comby2018, Blanchet2021}, and capable of identifying absolute configurations by comparison with theoretical calculations \cite{Powis2008c}. These properties originate from the fact that PECD mainly arises from the scattering of photoelectrons in the chiral molecular potential.

We interpret our experimental results by additionally advancing the theoretical methodology for PECD calculations. Whereas such calculations have previously mainly been performed for static molecules in the electronic ground state, we have developed the methods to calculate transient PECDs along photochemical reaction pathways. Whereas previous single-photon-PECD calculations relied on the continuum multiple scattering model using the X-Alpha local-exchange potential (CMS-X-Alpha) \cite{Powis2008,Powis2014} and density-functional-theory (DFT) \cite{Stener2004} methods, here we employ accurate quantum calculations of electron-molecular scattering to predict the PECD effects based on single-center partial wave expansion (Schwinger iterative method) \cite{Natalense1999, Gianturco1994}. We have additionally developed the formalism to include photoselection and molecular alignment into the calculations of PECD. The good agreement between calculated and observed PECDs deepens the understanding of the molecular quantum-scattering mechanisms underlying the observed PECD effects.

\begin{figure}[ht]
\centering
\includegraphics[width=\textwidth]{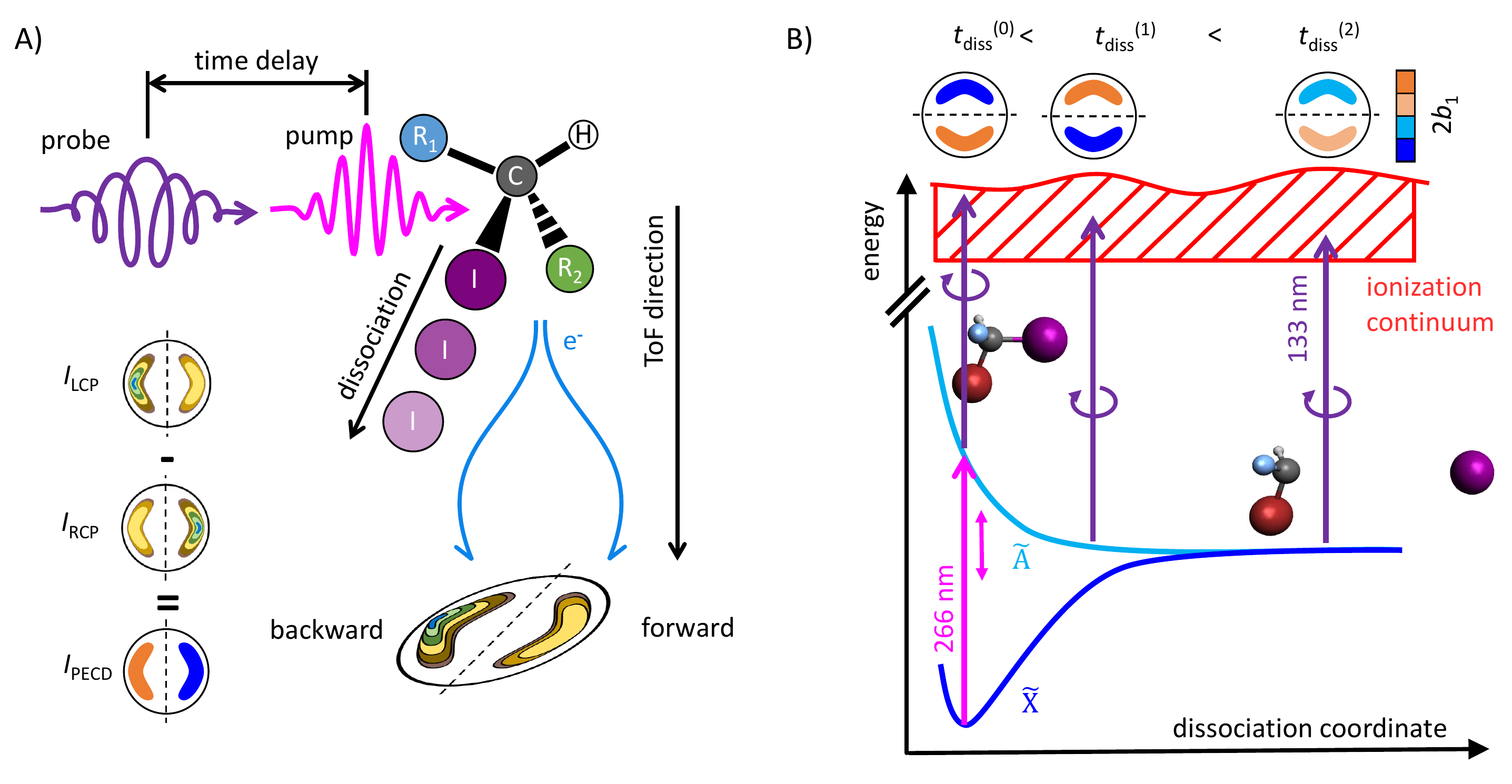}
\caption{{\bf Concept of VUV-TRPECD}. A) A pump pulse excites the system of interest and thereby initiates a photochemical reaction. A circularly polarized VUV probe pulse ionizes the photoexcited molecule and maps its time-dependent chirality onto a forward-backward asymmetry of the emitted photoelectrons with respect to the probe-pulse propagation direction. B) Illustration of time-resolved PECD in the photodissociation of CHBrFI.}
\label{fig:Figure1}
\end{figure}

The concept of VUV-TRPECD is illustrated in Fig.~\ref{fig:Figure1}. We investigate the time evolution of molecular chirality on the natural, femtosecond time scale of photochemical reactions by combining a source of ultrashort circularly polarized VUV pulses with photoelectron velocity-map imaging in a pump-probe scheme. As an example, we study the photodissociation reaction of chiral molecules excited by an ultrashort linearly polarized laser pulse and follow changes in the forward-backward asymmetry of the photoelectron angular distribution (PAD) represented by the sign and magnitude of the $2b_1$ coefficient (see the Supplementary Material SM, Section S2), which is a measure of the chirality, via a femtosecond circularly polarized laser pulse in the VUV range. For this purpose, we selected two molecules with complementary characteristics, bromofluoroiodomethane (CHBrFI) and 2-iodobutane (C$_4$H$_9$I). In both cases, the \ce{C-I} bond is selectively dissociated following one-photon absorption at 266~nm. The subsequent dynamics is tracked by a circularly polarized laser pulse centered at 133~nm generated with a new table-top source described in the SM, Section S1 and Ref. \cite{Svoboda2022}. Our measurements reveal intriguing changes in the PECD signal on the time scale of the bond-breaking dynamics and additionally reveal the chirality of the radical products. A particular property of this scheme is the selectivity of the probe, which discriminates between the fragments, revealing the chirality changes in the nascent molecular radical.

\section*{Results}

\begin{figure}[ht]
\centering
\includegraphics[width=\linewidth]{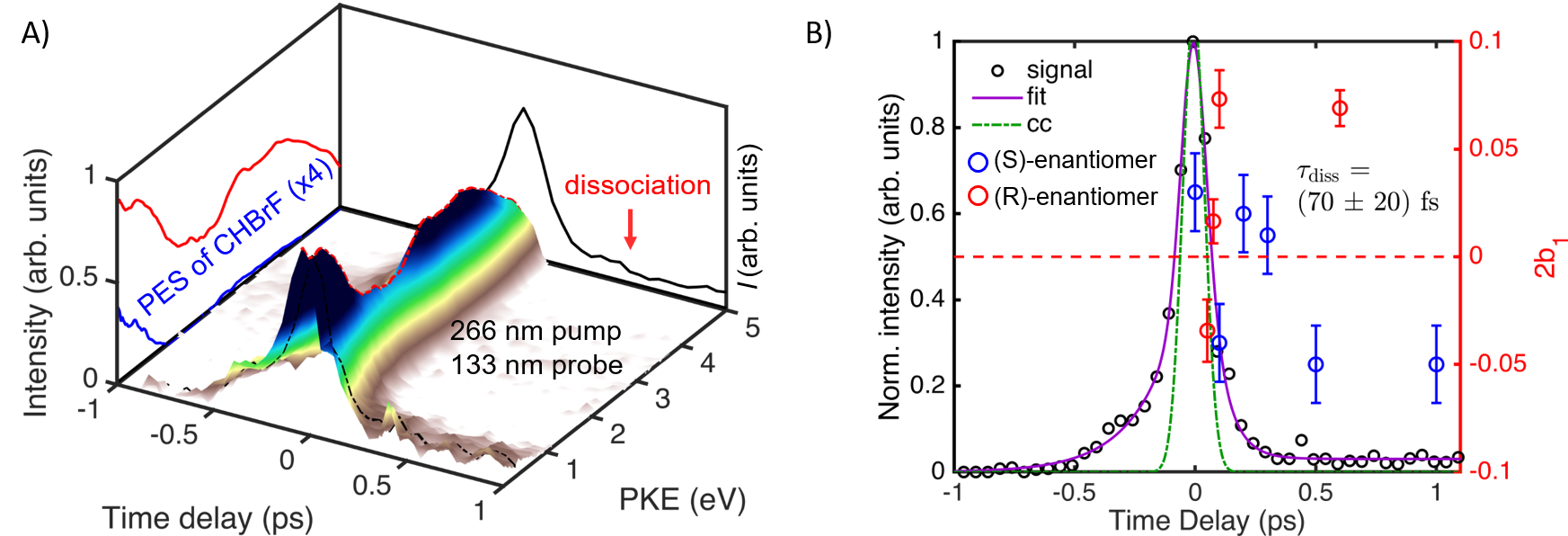}
\caption{{\bf Probing time-dependent chirality in the photodissociation of CHBrFI}. A) Time-resolved photoelectron spectra obtained with pump and probe laser pulses centered at 266~nm and 133~nm, respectively (positive delays) and vice versa (negative delays). Photoelectron spectrum of CHBrF radical (blue solid line) is obtained by averaging the time-resolved spectra between 0.3~ps and 1~ps. B) Photoelectron signal integrated over kinetic energy from 0~eV to 1.5~eV (left \textit{y}-axis - black) and time-dependent PECD of the photodissociation of both enantiomers of CHBrFI (right \textit{y}-axis - red). CC stands for the cross-correlation signal (green dash-dotted line).}
\label{fig:Figure2}
\end{figure}

The time-resolved photoelectron spectrum (TRPES) of the photodissociation of CHBrFI (vertical ionization energy is 9.86~eV \cite{Novak2002}) is presented in Fig.~\ref{fig:Figure2}A. The dynamics on the positive-delay side (Fig.~\ref{fig:Figure1}B) corresponds to the CHBrFI molecule photodissociating on the $\tilde{\text{A}}$-state potential energy surface into a CHBrF$^{\bullet}$ radical and an iodine atom. The observed spectra are dominated by a time-zero feature corresponding to resonance-enhanced (1+1$^{\prime}$)-two-photon ionization. On the positive delay side, the photoelectron signal decays rapidly within the first 300~fs and remains constant afterwards. This constant signal forms a broad spectral feature spanning a kinetic-energy range from 0~eV to $\sim$1~eV, which is assigned to the CHBrF$^{\bullet}$ radical based on its calculated adiabatic ionization energy of 8.32~eV \cite{He2009}. The iodine atom cannot be ionized by one photon of 133~nm (9.3~eV) because its ionization energy is too high (10.45~eV\cite{Imre1980}). The time-resolved map of the associated asymmetry parameters $\beta_2$ is shown in the Supplementary Material (Fig. S2).

The dissociation time is obtained from the signal in the photoelectron kinetic energy range between 0~eV and 1.5~eV using a mono-exponential fit of the form $A \cdot \exp(-t/\tau_{\mathrm{diss}}) + B$ convoluted with a Gaussian cross-correlation (cc, Fig.~\ref{fig:Figure2}B). The extracted dissociation time is $\tau_{\mathrm{diss}} = (70 \pm 20)$~fs (see Fig.~\ref{fig:Figure2}B). This value agrees well with a theoretical estimate of $\tau_{\mathrm{diss}} ^{(\mathrm{calc})} = 77$~fs, obtained using a classical propagation on a non-relativistic \textit{ab-initio} surface, which assumed that the dissociation is completed at $r(\ce{C-I}) = 4.5$~\AA.

The negative time-delay side represents the photo-excitation to Rydberg states by the 133-nm pulse. The evolution on this Rydberg-state manifold is followed by one-photon ionization with the 266-nm pulse. On the negative-delay side, the signal decays to zero in about 0.5~ps. This decay is much faster than the expected lifetime of the Rydberg states and is therefore attributed to the photoexcited wave packet leaving the configurational space from where one-photon ionization by the 266-nm pulse is possible. The signal in the energy range between 2.5~eV and 5.0~eV (see Fig.~\ref{fig:Figure2}A) is well represented by a mono-exponential decay with a lifetime of $\tau_{\mathrm{Rydberg}} = (97 \pm 6)$~fs.

After discussion of the TRPES results, we now turn to the TRPECD results. Enantiomerically pure ($\sim 90$~\%~ee) samples of CHBrFI have been obtained by chiral chromatographic separation of racemic samples as described in the SM Section S4. The pump-probe measurements were performed by using circularly polarized 133-nm pulses to ionize the molecule following excitation by a linearly polarized 266-nm pulse. The PECD images are formed by subtracting inverted photoelectron images measured with left- and right-circularly polarized 133-nm pulses. These images are inverted using a modified version of the Onion-peeling method, see SM Section S2 for more details. The resulting difference image contains only the chiral contribution to the photoelectron angular distribution (PAD). The chiral signal integrated over the photoelectron kinetic energy range corresponding to the nascent CHBrF fragment (0~eV to 1.5~eV) is determined by a fit of the form $2b_1 \cos\theta$ to the PECD image, where $\theta$ is the photoelectron emission angle with respect to the propagation direction of the circularly polarized light. 

The TRPECD observed during the photodissociation of CHBrFI is shown in Fig.~\ref{fig:Figure2}B.  Higher-order Legendre contributions to the PECD \cite{Powis2008b} are found to be negligible ($b_3 = (0.00 \pm 0.01)$) for both enantiomers and all measured time delays. 
In the case of the (S)-enantiomer, the PECD value at 0~fs time delay amounts to $2b_1 = (0.03 \pm 0.01)$, followed by a sign inversion to $2b_1 = (-0.04 \pm 0.01)$ at 100~fs delay, another sign inversion to delays of 200~fs and 300~fs, and a third inversion of the PECD up to delays of 500~fs.
Due to limitations in the quantity of enantiopure samples, the measurements on the (R)-enantiomer were focused on the early time delays, which confirmed the inversion of the PECD to lie between 50~fs and 80~fs, and a late time delay, which confirmed the chirality of the photoproduct radical. Overall, the measured PECD values display the inversion symmetry that is expected upon exchange of the enantiomers. 

\begin{figure}[ht]
\centering
\includegraphics[width=\linewidth]{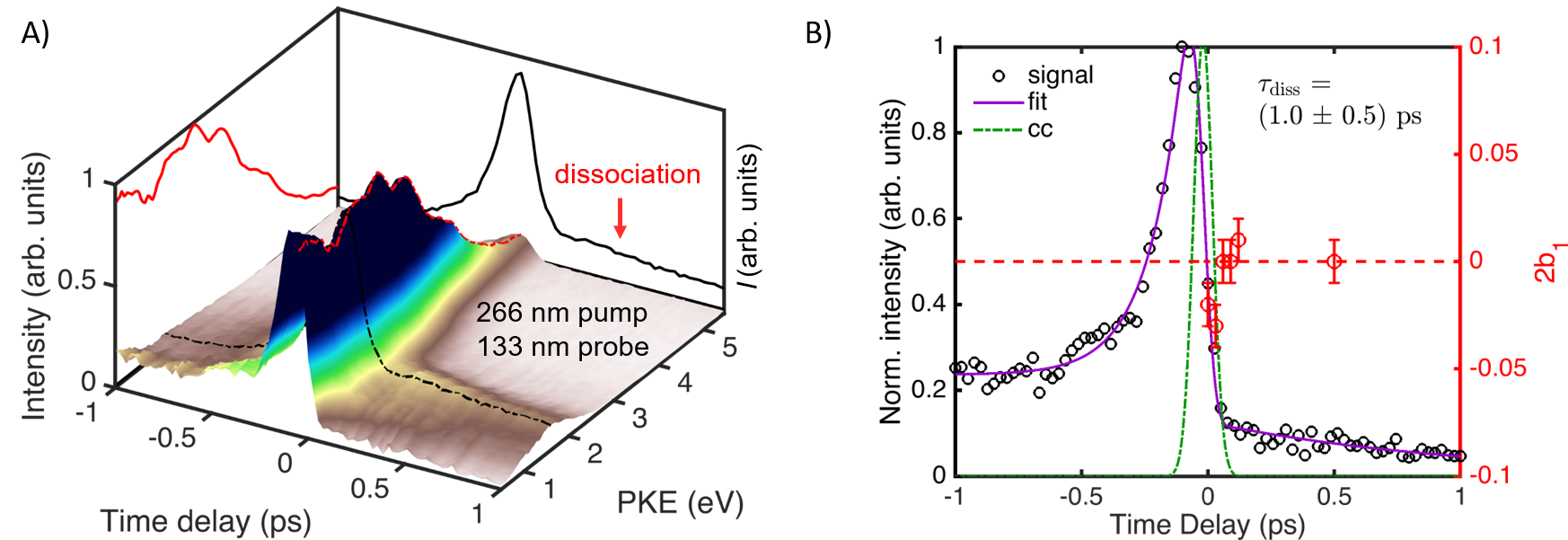}
\caption{{\bf Probing time-dependent chirality in the photodissociation of 2-iodobutane}. A) Time-resolved photoelectron spectra obtained with pump and probe laser pulses centered at 266~nm and 133~nm, respectively (positive delays) and vice versa (negative delays). B) Photoelectron signal integrated from  0.7~eV to 2.5~eV kinetic energy (left \textit{y}-axis - black) and time-dependent PECD of the photodissociation of 2-iodobutane (right \textit{y}-axis - red). CC stands for the cross-correlation signal (green dash dotted line).
}
\label{fig:Figure3}
\end{figure}

The generality of our VUV-TRPECD method is demonstrated through its application to another molecule, 2-iodobutane. This molecule is structurally more complex because the atomic substituents (Br and F) at the chiral center are replaced with methyl and ethyl groups that have internal degrees of freedom (internal rotations and vibrations). As in the case of CHBrFI, we study the selective \ce{C-I} bond dissociation induced by a pump pulse centered at 266~nm. 

The TRPES is shown in Fig.~\ref{fig:Figure3}A. Positive time delays correspond to the dissociation of the \ce{C-I} bond. The dissociation produces a 2-butyl radical and an iodine atom, whereby the latter does not contribute to the photoelectron spectrum for the same energetic reasons as stated above. Since the vertical ionization energy of the 2-butyl radical amounts to 7.6~eV \cite{Schultz1984}, the corresponding photoelectron band should lie at $\mathrm{PKE} = 1.7$~eV. This value agrees well with the measured transient spectrum. The signal at negative time delays corresponds to an excitation to high-lying Rydberg states \cite{Park2001}. The time-resolved map of the associated asymmetry $\beta_2$ parameters is shown in the Supplementary Material (Fig. S2).

The observed dynamics are analyzed by a fit to the spectrum in the energy range between 0.7~eV and 2.5~eV. The fitting results are shown in Fig.~\ref{fig:Figure3}B, where a mono-exponential decay of the form $A \cdot \exp(-t/\tau_{\mathrm{diss}})$, convoluted with a Gaussian cross-correlation (cc), is fitted to the positive-delay side of the TRPES. The dissociation is completed within the temporal overlap of the pump and probe pulses. The photoelectron signal originating from the 2-butyl radical then further decays with a time constant $\tau_{\mathrm{diss}} = (1.0 \pm 0.5)$~ps, which is assigned to vibrational relaxation \cite{Corrales2014}. A fit to the signal on the negative-delay side yields a time constant $\tau_{\mathrm{Rydberg}} = (140 \pm 20)$~fs for the wave packet dynamics created in the Rydberg states.

A distinctive feature of 2-iodobutane is that the ground state can be one-photon ionized by 133-nm radiation. This allowed us to measure a static PECD from the unexcited molecule as well as a transient PECD from the dissociating state of the molecule. The static PECD for the ground state amounts to $2b_1 = (-0.11 \pm 0.02)$, represented by a green diamond in Fig.~\ref{fig:Figure4}B, which additionally shows the PECD image. The PECD at the temporal overlap of the pump and probe pulses is much smaller in magnitude ($2b_1=-0.02 \pm 0.01$). Subsequently, the PECD increases to reach a maximum of $2b_1= -0.03 \pm 0.01$ at a delay of 30~fs and decays to zero by 60~fs, remaining close to zero for all later time delays.

\subsection*{Discussion}

\begin{figure}[ht]
\centering
\includegraphics[width=\linewidth]{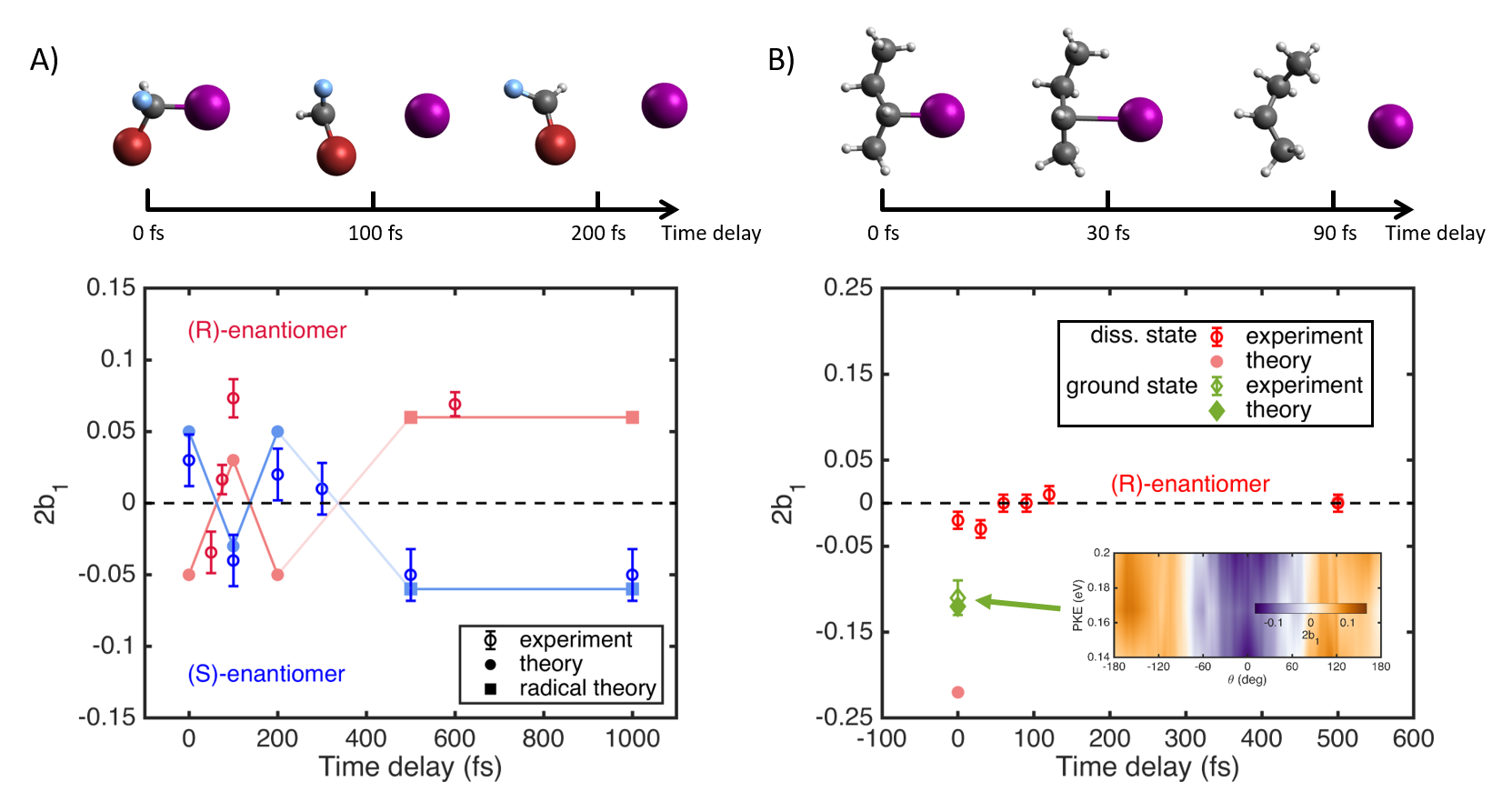}
\caption{{\bf Comparison of calculated and measured PECDs and their interpretation}. A) TRPECD during the photodissociation of CHBrFI. The lines connecting the calculated points are meant as a guide to the eyes only. B) TRPECD (circles) of the photodissociation and ground state PECD (diamonds) of the (R)-enantiomer of 2-iodobutane. The inset shows a PECD image in polar coordinates obtained from the ground state of 2-iodobutane. Error bars represent $\pm2\sigma$ error from the non-linear fit. Details are given in the main text.}
\label{fig:Figure4}
\end{figure}

To rationalize the observed TRPECD signals, we have developed a new method to calculate the PECD effect on the basis of accurate electron-molecule scattering methods. Details about this work are described in the SM Section S3. The PECD associated with the electronically excited state of CHBrFI prepared in the experiment was calculated as follows. The excited-state wavefunction at the equilibrium geometry of the electronic ground state was obtained through a state-specific complete-active-space self-consistent-field  (CASSCF) calculation using the Molpro package \cite{Werner2011, Molpro2015}. The ePolyScat package \cite{Gianturco1994, Natalense1999} was then used to obtain the photoionization matrix elements from which the chiral-sensitive PAD coefficient $b_1$ was extracted in a subsequent step. 

We first discuss the TRPECDs measured in the case of CHBrFI. The experimental data is reproduced in Fig.~\ref{fig:Figure4}A as open circles, whereas the calculated results are shown as filled circles. 
Most importantly, the calculations reproduce the sign alternation of the PECD observed in the experiment, as well as the order of magnitude of the transient PECD effects. Specifically, the sign changes from $t=0$ to $t=100$~fs, observed for both enantiomers, and the additional inversions from $t=100$~fs to $t=200$~fs and from $t=200$~fs to $t=500$~fs, observed for the (S)-enantiomer, are all reproduced.
All calculated PECDs shown in this article were obtained for isotropic axis distributions. The effect of photoselection does not change qualitatively the results, as discussed in the SM Section S3. Calculations for delays longer than $t=200$~fs could not be numerically converged because they correspond to \ce{C-I} bond lengths larger than 4.65~\AA, which cannot be efficiently described within the single-center-expansion used in ePolyScat. Instead, the calculated PECD values given for $t=500$~fs and $t=1000$~fs (blue squares) were obtained from a calculation on the equilibrium geometry of the isolated CHBrF$^{\bullet}$ radical in its electronic ground state. At this delay, the CHBrF$^{\bullet}$ radical can be safely assumed to behave as an isolated entity. In this case, we even obtain a quantitative agreement with the measured values obtained from both enantiomers.

The measured TRPECD signals of 2-iodobutane are reproduced in Fig.~\ref{fig:Figure4}B in red. At time zero, the experimental values of both the ground and the first excited states are shown (green diamond and red circle, respectively). Both measured PECDs are negative, but the ground-state PECD is significantly larger in absolute value. In the case of the electronic ground state, the experimental and calculated PECDs agree well. In the case of the electronically excited state at temporal overlap of pump and probe pulses, the measured PECD is significantly smaller than the calculated value. This discrepancy might be caused by the contribution of a second photoionization channel, in which the 133-nm photon is absorbed first and the 266-nm photon second. This channel is not included in our calculations. Attempts to calculate PECDs for longer time delays, i.e. larger \ce{C-I} separations, were not successful because the ePolyScat calculations did not converge. However, the PECD of the 2-butyl radical in its electronic ground state could be obtained. It is shown as the filled red circle at a delay of 500~fs, which agrees well with the vanishing measured PECD.

The most remarkable feature of the measured TRPECDs is their different behavior in the two molecules. Whereas the TRPECD of CHBrFI shows pronounced variations as a function of time, including at least three inversions of its sign, the TRPECD of 2-iodobutane converges rapidly to zero. These different dynamics can be explained in terms of the structural dynamics of the nascent radicals. Whereas the CHBrF$^{\bullet}$ radical has a single low frequency vibrational mode, i.e. the umbrella motion, the 2-butyl radical has many soft modes, including internal rotations and the H-wagging mode.

\begin{figure}[ht]
\centering
\includegraphics[width=\linewidth]{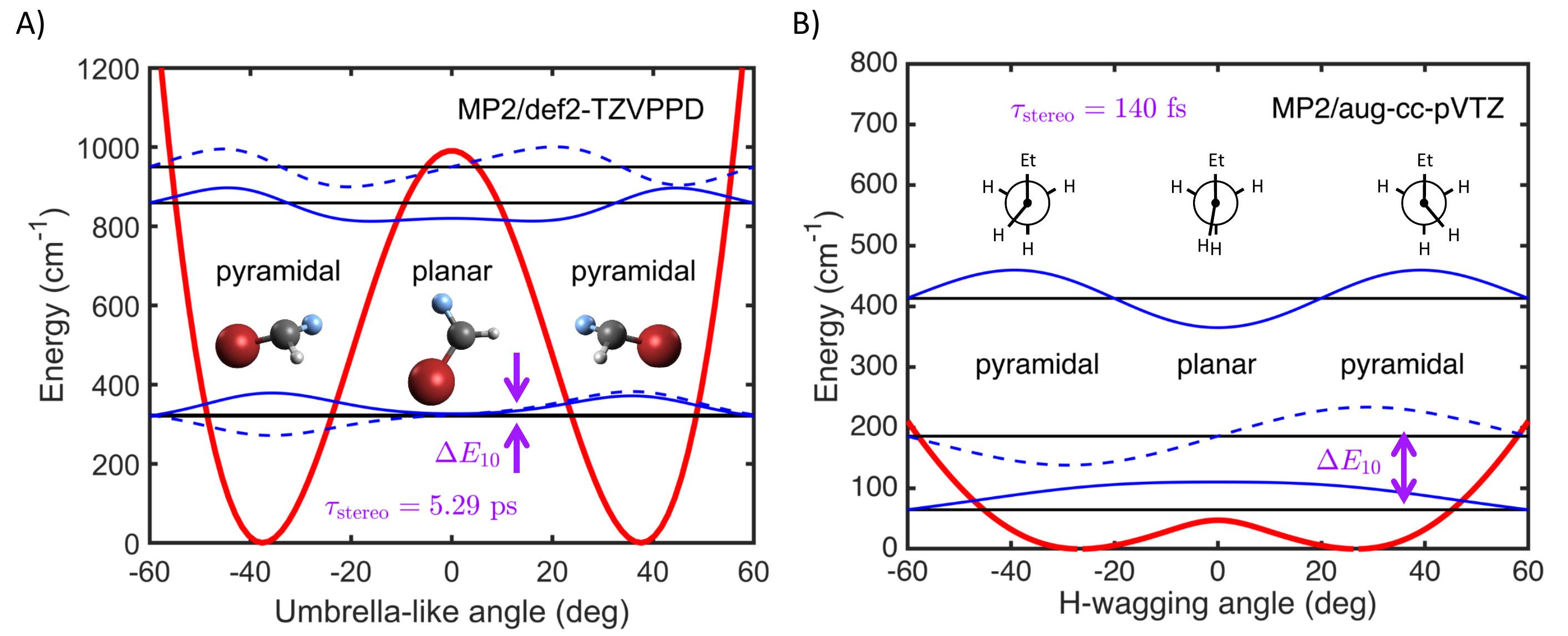}
\caption{{\bf Illustration of the origin of photoproduct chirality.} A) Potential energy surface of the CHBrF radical along the umbrella mode and its lowest lying vibrational eigenstates. B) Potential energy surface of the 2-iodobutane radical along the hydrogen-wagging coordinate and its lowest lying vibrational eigenstates. Blue solid and dashed lines discern the lower and upper vibrational states in tunneling doublets.}
\label{fig:Figure5}
\end{figure}

To understand these differences, we discuss the topology of the potential energy surfaces of the two radicals along the coordinate that interconverts the two enantiomers, i.e. the umbrella mode in CHBrFI and the H-wagging mode in the 2-butyl radical (see Fig.~\ref{fig:Figure5}). 
The one-dimensional cut of the potential energy surface of the CHBrF$^{\bullet}$ radical along the umbrella mode exhibits a double well structure as shown in Fig. \ref{fig:Figure5}A. The two minima correspond to two enantiomeric pyramidal geometries of the radical, whereas the local maximum of the potential corresponds to a planar achiral geometry. The characteristic time scale for the transformation of one enantiomeric structure into the other, which occurs by tunneling when the total energy lies below the barrier, is also known as the stereomutation time $\tau_{\mathrm{stereo}} = h/(2\Delta E_{10})$ \cite{Quack1986}, where $h$ is the Planck constant and $\Delta E_{10}$ is the energy splitting between the lowest and the first excited vibrational states of the inversion mode. The vibrational energy levels, calculated using a one-dimensional discrete-variable-representation (DVR) approach \cite{Colbert1992} along the umbrella-like coordinate, are shown as horizontal lines. The stereomutation time for the CHBrF$^{\bullet}$ radical in its lowest pair of eigenstates is $\tau_{\mathrm{stereo}} = 5.29$~ps. Because the dissociation time is much shorter than the stereomutation time, the CHBrF$^{\bullet}$ radical may be formed in one of the two possible pyramidal forms, which corresponds to a coherent superposition of at least the two lowest states of the umbrella mode. Our observation of a non-vanishing PECD for pump-probe delays at which dissociation is completed (0.5-1.0~ps) is consistent with a long stereomutation time, i.e. the formation of a photoproduct that remains chiral on the time scale of our measurements.

\begin{figure}[ht]
\centering
\includegraphics[width=0.5\textwidth]{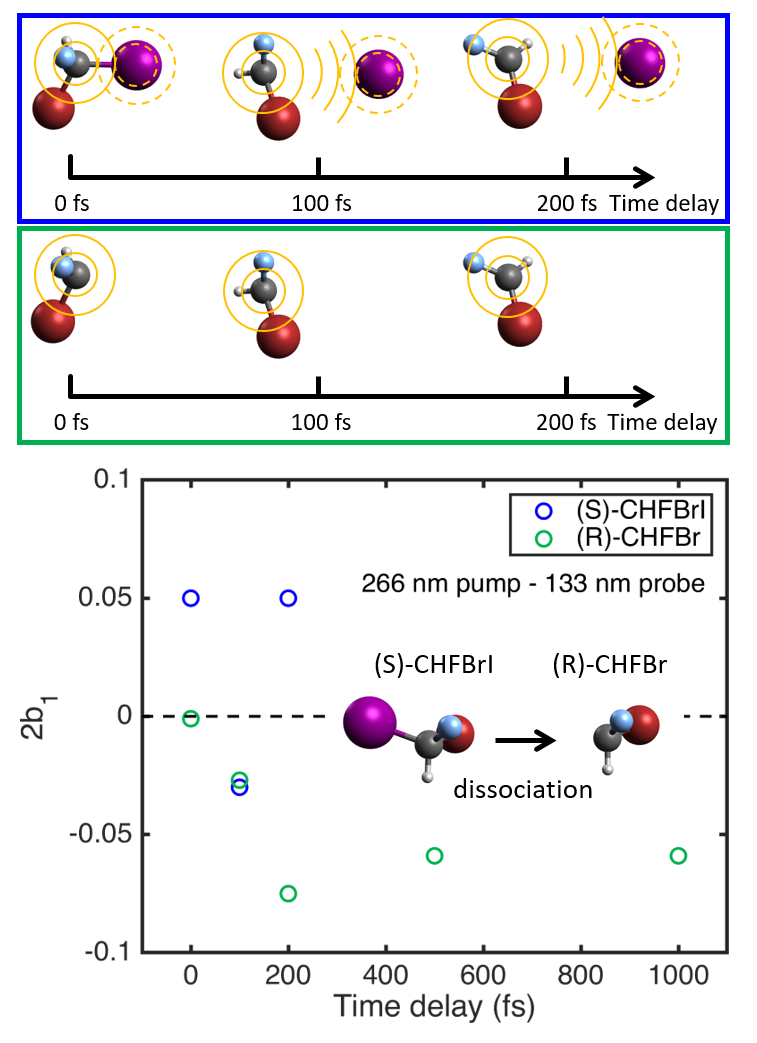}
\caption{{\bf Effect of photoelectron scattering off the iodine atom (yellow circles) on TRPECD.} Calculations of the PECD of the dissociating molecule (blue) and the CHBrF$^{\bullet}$ radical using the same atomic coordinates, but removing the iodine atom (green).}
\label{fig:Figure6}
\end{figure}

Figure~\ref{fig:Figure5}B shows a one-dimensional cut of the potential energy surface of the 2-butyl radical along the H-wagging coordinate. This potential also has a double well structure, but in this case, the barrier height is much lower. This leads to a situation where the lowest vibrational state lies above the barrier and results in an effectively achiral equilibrium structure. This explains both the vanishing calculated PECD for the 2-butyl radical and the rapid decay of the measured PECD to zero.

The most intriguing aspect of our TRPECD results is the observation of multiple sign changes of the PECD effect during the photodissociation of CHBrFI. The time delays $t=100$~fs and $t=200$~fs correspond to \ce{C-I} distances of more than 4~\AA\xspace over which the potential energy surfaces are flat. Since the probe pulse cannot ionize atomic iodine, ionization happens exclusively from the CHBrF$^{\bullet}$ radical at both delays. The sign changes in the PECD effect could therefore originate from inversions of the CHBrF$^{\bullet}$ radical along the umbrella mode. To verify this hypothesis, we have repeated the PECD calculations shown in Fig.~\ref{fig:Figure4}A by using the same molecular geometries but removing the iodine atom, repeating the electronic structure and finally the photoionization calculations to obtain the PECD of the CHBrF$^{\bullet}$ fragment. The results, shown as green circles in Fig.~\ref{fig:Figure6}, are negative for all time delays, in contrast to the PECD calculations including the iodine atom (blue circles). This shows that the sign inversions of the PECD do not originate from the inversions of the CHBrF$^{\bullet}$ radical. They must therefore be caused by the presence of the dissociating iodine atom. Since the latter is separated by $>4$~\AA~from the CHBrF radical at delays $>100$~fs, and has therefore no effect on the electronic structure of the radical, the inversions of the PECD must originate from scattering of the photoelectrons originating from the CHBrF$^{\bullet}$ radical on the neutral iodine atom. The dramatic effect of electron scattering from the iodine atom on the PECD is particularly apparent by comparing PECD values at pump-probe delays $t=100$~fs and $t=200$~fs. At the earlier time delay, the PECD values are very similar, whereas they are almost opposite at the later delay. Remarkably, the sensitivity of this PECD effect extends well beyond that of photoelectron spectroscopy. Whereas the photoelectron spectrum remains essentially unchanged after $t=100$~fs (Fig.~\ref{fig:Figure2}) because the dissociating wave packet has reached the flat part of the potential energy surfaces, the PECD is still highly sensitive to the presence of the dissociating atom, to the level that it changes sign several times. These results highlight the extreme sensitivity of TRPECD to the structural evolution of the dynamically evolving molecular environment, beyond the simpler energetic aspects probed by TRPES. 

In this work, we have demonstrated a new general technique for probing chirality changes during a chemical reaction. We have revealed the potential of VUV-TRPECD by time-resolving the changes in molecular chirality introduced by the selective \ce{C-I} bond dissociation of two chiral molecules with complementary characteristics. Our results were interpreted by comparison with a new theoretical method for calculating PECDs based on accurate electron-molecule scattering calculations.

In the case of CHBrFI, we have observed a pronounced variation including sign changes of the PECD during the photodissociation reaction. These inversions of the PECD have been assigned to photoelectron scattering off the dissociating iodine atom on the basis of our calculations. The measured PECD at long time delays differs from zero and agrees with the calculated PECD of the chiral equilibrium geometry of the CHBrF$^{\bullet}$ radical. In the case of 2-iodobutane, we have measured the PECD of both the excited and unexcited molecules. Our measurements have revealed a rapid decay of the PECD to zero within 60~fs. This observation is explained in terms of a rapid racemization of the nascent 2-butyl radical rationalized by the achiral equilibrium geometry of the 2-butyl radical.

Overall, we have demonstrated that VUV-TRPECD opens a general route to studies of dynamical chirality during chemical reactions. This approach is sufficiently sensitive to be applicable to molecules in the gas phase. This general method gives access to the chirality of complete photochemical reaction pathways from photoexcitation through conical intersections to the final products, without the restrictions that are imposed by photon energies in the VIS/UV range. Hence, this technique has the potential to drive major new developments in the understanding of the dynamic evolution of molecular chirality in chemical reactions and its role in chiral recognition and enantioselective molecular processes in general.

\section*{Materials and Methods}
\subsection*{Experiment}

All experiments have been performed using a two-color pump-probe scheme utilizing low-order-harmonic generation in a semi-infinite gas cell and a velocity-map-imaging (VMI) spectrometer \cite{Svoboda2017}. A Ti:Sa regenerative amplifier operating at 1~kHz delivered 2.0~mJ pulses centered at 800~nm with 28-fs pulse duration. The input beam was divided into two arms using an 80:20 beam splitter. The reflected beam was frequency-doubled using a 300~$\upmu$m thick $\beta$-barium borate crystal to obtain 570~$\upmu$J pulses at 400~nm ($\sim 35$~fs). The second harmonic beam was separated from the fundamental beam by reflections on four dichroic mirrors. The polarization of the beam was controlled by a motorized rotational stage with a quarter-wave plate for 400~nm \cite{Svoboda2022}. The beam was focused by a $f = 500$~mm spherical mirror into a semi-infinite gas cell filled with Xe. The Xe pressure $p_{\mathrm{Xe}} = 10$~mbar was selected such that the third (133~nm $\sim$ 9.3~eV) harmonic of 400~nm was optimized. The transmitted beam was used for the third harmonic generation giving 1.2~$\upmu$J pulses of 266~nm ($\sim 70$~fs). Both beams were delivered into a vacuum chamber using their respective dichroic mirrors and then focused non-collinearly (with a crossing angle of less than 1$^{\circ}$) by two spherical mirrors ($f = 0.5$~m for the harmonic beam and $f = 1$~m for the 266~nm beam). All in-vacuum mirrors were motorized allowing for fine adjustment of the spatial overlap.

Samples (racemic or enantiomerically enriched) were evaporated and mixed with Ne as a carrier gas. The mixture was delivered into the vacuum chamber through a pulsed nozzle (orifice 150~$\upmu$m, 1~kHz repetition rate). The gas jet created by the nozzle was skimmed by a 500~$\upmu$m skimmer and propagated 10~cm downstream to reach the interaction region where it was ionized by the combined action of the laser beams. A cross-correlation between both pulses was determined by non-resonant two-color ionization (133~nm + 266~nm) of Xe. The cross-correlation function was taken into account during data analysis as an instrument-response function. Ejected photoelectrons were imaged using electrostatic lenses fulfilling VMI conditions, and a dual microchannel-plate detector in a Chevron configuration.

\subsection*{Theory}

Ground state geometries for both molecules are optimized with Orca program package on MP2/aug-cc-pVTZ level of theory. All molecular geometries for the first excited state needed for subsequent CASSCF and ePolyScat calculations are obtained from \textit{ab-initio} molecular dynamics calculations. The Born-Oppenheimer molecular dynamics (BOMD) on the first excited singlet state, calculated at the CIS/6-311G$^\ast$ level with 4 singlet states and no triplet states, is performed using the Q-Chem program package. The time step of the BOMD is 25~a.u. (0.6~fs) and the initial velocities are set to zero.

The electronic structure calculations are done in MOLPRO program package. Since ePolyScat only works with Hartree-Fock-type wavefunctions in which each orbital is either doubly occupied, singly occupied, or empty, we described the first excited state of the molecules using a simple CASSCF-type wavefunction using an active space consisting of two electrons in two orbitals, CAS(2,2), with only two electronic states with the ground-state weight 0 and the first-excited-state weight 1. We used cc-pVTZ basis set on all atoms except iodine which was described with all electron 6-311G$^\ast$ basis set. As a result, the first excited state was described as (...)$^2$(HOMO)$^1$(LUMO)$^1$.

Photoionization matrix elements are calculated with ePolyScat program package \cite{Gianturco1994, Natalense1999}, which calculates the matrix elements by solving the corresponding quantum-mechanical scattering problem in a single-center partial wave expansion using the Schwinger variational principle. For each time delay, two calculations are done in which either the HOMO or LUMO orbital is ionized and the other orbital is left singly occupied. From the calculated photoionization matrix elements, $b_1$ coefficients are calculated in a separate custom-built routine and HOMO and LUMO contributions are averaged to get the final $b_1$ coefficient. 

\section*{Supplementary materials}
Section S1. Polarization measurements \\
Section S2. PECD measurements \\
Section S3. PECD calculation \\
Section S4. Synthesis of CHBrFI \\
Section S5. Synthesis of 2-iodobutane \\
Section S6. Photoelectron angular distributions\\

\section*{Acknowledgements}

The authors thank Andreas Schneider, Mario Seiler, and Andres Laso for their contributions to the construction and improvement of the experiment. J.S. acknowledges help from and discussion with co-worker Manuela Meister, Vanessa Galati, and Stole Manov, the latter also provided a first batch of enantiopure 2-iodobutane. V.S. and H.J.W. acknowledge a fruitful discussion with Prof.~Stephen Pratt. J.S. received financial support from ZHAW under grant AF-2018 and AF-2019. We gratefully acknowledge funding from ETH Zurich and the Swiss National Science Foundation through the NCCR-MUST and grant 20021\_172946.

\section*{Author contributions statement}

H.J.W. proposed the experiments. V.S. and N.B.R. designed and assembled the experiment. V.S. conducted the experiments, analysed the data, and performed the theoretical calculations. D.B. and H.H. developed the codes for the PECD calculations. M.D.J.W. helped with some experiments. V.S., B.S. and J.S. prepared the enantiopure samples of CHBrFI. M.O. and D.Z. prepared the enantiopure sample of 2-iodobutane. The manuscript was written by V.S. and H.J.W. All authors reviewed the manuscript.

\section*{Additional information}

\textbf{Competing interests} The authors declare that they have no competing interests.

\noindent
\textbf{Data and material availability} All data needed to evaluate the conclusions in the paper are present in the paper and/or the Supplementary Materials.

\noindent
\textbf{Correspondence} Correspondence and request for materials should be addressed to H.J. W\"orner (email: hwoerner@ethz.ch) or V. Svoboda (email:vit.svoboda@phys.chem.ethz.ch).

\clearpage

\bibliography{References}

\begin{thebibliography}{10}
\urlstyle{rm}
\expandafter\ifx\csname url\endcsname\relax
  \def\url#1{\texttt{#1}}\fi
\expandafter\ifx\csname urlprefix\endcsname\relax\def\urlprefix{URL }\fi
\expandafter\ifx\csname doiprefix\endcsname\relax\def\doiprefix{DOI: }\fi
\providecommand{\bibinfo}[2]{#2}
\providecommand{\eprint}[2][]{\url{#2}}

\bibitem{Meyer-Ilse2012}
\bibinfo{author}{Meyer-Ilse, J.}, \bibinfo{author}{Akimov, D.} \&
  \bibinfo{author}{Dietzek, B.}
\newblock \bibinfo{journal}{\bibinfo{title}{Ultrafast circular dichroism study
  of the ring opening of 7-dehydrocholesterol}}.
\newblock {\emph{\JournalTitle{Journal of Physical Chemistry Letters}}}
  \textbf{\bibinfo{volume}{3}}, \bibinfo{pages}{182--185},
  \doiprefix\url{10.1021/jz2014659} (\bibinfo{year}{2012}).

\bibitem{Oppermann2019}
\bibinfo{author}{Oppermann, M.} \emph{et~al.}
\newblock \bibinfo{journal}{\bibinfo{title}{Ultrafast broadband circular
  dichroism in the deep ultraviolet}}.
\newblock {\emph{\JournalTitle{Optica}}} \textbf{\bibinfo{volume}{6}},
  \bibinfo{pages}{56--60}, \doiprefix\url{10.1364/OPTICA.6.000056}
  (\bibinfo{year}{2019}).

\bibitem{Patterson2013}
\bibinfo{author}{Patterson, D.}, \bibinfo{author}{Schnell, M.} \&
  \bibinfo{author}{Doyle, J.~M.}
\newblock \bibinfo{journal}{\bibinfo{title}{Enantiomer-specific detection of
  chiral molecules via microwave spectroscopy}}.
\newblock {\emph{\JournalTitle{Nature}}} \textbf{\bibinfo{volume}{497}},
  \bibinfo{pages}{475--477}, \doiprefix\url{10.1038/nature12150}
  (\bibinfo{year}{2013}).

\bibitem{Hirota2012}
\bibinfo{author}{Hirota, E.}
\newblock \bibinfo{journal}{\bibinfo{title}{Triple resonance for a three-level
  system of a chiral molecule}}.
\newblock {\emph{\JournalTitle{Proceedings of the Japan Academy, Series B}}}
  \textbf{\bibinfo{volume}{88}}, \bibinfo{pages}{120--128},
  \doiprefix\url{10.2183/pjab.88.120} (\bibinfo{year}{2012}).

\bibitem{Pitzer2013}
\bibinfo{author}{Pitzer, M.} \emph{et~al.}
\newblock \bibinfo{journal}{\bibinfo{title}{Direct determination of absolute
  molecular stereochemistry in gas phase by {C}oulomb explosion imaging}}.
\newblock {\emph{\JournalTitle{Science}}} \textbf{\bibinfo{volume}{341}},
  \bibinfo{pages}{1096--1100}, \doiprefix\url{10.1126/science.1240362}
  (\bibinfo{year}{2013}).

\bibitem{Herwig2013}
\bibinfo{author}{Herwig, P.} \emph{et~al.}
\newblock \bibinfo{journal}{\bibinfo{title}{Imaging the absolute configuration
  of a chiral epoxide in the gas phase}}.
\newblock {\emph{\JournalTitle{Science}}} \textbf{\bibinfo{volume}{342}},
  \bibinfo{pages}{1084--1086}, \doiprefix\url{10.1126/science.1246549}
  (\bibinfo{year}{2013}).

\bibitem{Cireasa2015}
\bibinfo{author}{Cireasa, R.} \emph{et~al.}
\newblock \bibinfo{journal}{\bibinfo{title}{Probing molecular chirality on a
  sub-femtosecond~timescale}}.
\newblock {\emph{\JournalTitle{Nature Physics}}} \textbf{\bibinfo{volume}{11}},
  \bibinfo{pages}{654--658}, \doiprefix\url{10.1038/nphys3369}
  (\bibinfo{year}{2015}).

\bibitem{Harada2018}
\bibinfo{author}{Harada, Y.}, \bibinfo{author}{Haraguchi, E.},
  \bibinfo{author}{Kaneshima, K.} \& \bibinfo{author}{Sekikawa, T.}
\newblock \bibinfo{journal}{\bibinfo{title}{Circular dichroism in high-order
  harmonic generation from chiral molecules}}.
\newblock {\emph{\JournalTitle{Physical Review A}}}
  \textbf{\bibinfo{volume}{98}}, \bibinfo{pages}{021401},
  \doiprefix\url{10.1103/physreva.98.021401} (\bibinfo{year}{2018}).

\bibitem{Baykusheva2018}
\bibinfo{author}{Baykusheva, D.} \& \bibinfo{author}{W\"orner, H.~J.}
\newblock \bibinfo{journal}{\bibinfo{title}{Chiral discrimination through
  bielliptical high-harmonic spectroscopy}}.
\newblock {\emph{\JournalTitle{Phys. Rev. X}}} \textbf{\bibinfo{volume}{8}},
  \bibinfo{pages}{031060}, \doiprefix\url{10.1103/PhysRevX.8.031060}
  (\bibinfo{year}{2018}).

\bibitem{Ritchie1976}
\bibinfo{author}{Ritchie, B.}
\newblock \bibinfo{journal}{\bibinfo{title}{Theory of the angular distribution
  of photoelectrons ejected from optically active molecules and molecular
  negative ions}}.
\newblock {\emph{\JournalTitle{Physical Review A}}}
  \textbf{\bibinfo{volume}{13}}, \bibinfo{pages}{1411--1415},
  \doiprefix\url{10.1103/physreva.13.1411} (\bibinfo{year}{1976}).

\bibitem{Powis2000}
\bibinfo{author}{Powis, I.}
\newblock \bibinfo{journal}{\bibinfo{title}{Photoelectron circular dichroism of
  the randomly oriented chiral molecules glyceraldehyde and lactic acid}}.
\newblock {\emph{\JournalTitle{Journal of Chemical Physics}}}
  \textbf{\bibinfo{volume}{112}}, \bibinfo{pages}{301--310},
  \doiprefix\url{10.1063/1.480581} (\bibinfo{year}{2000}).

\bibitem{Bowering2001}
\bibinfo{author}{B\"owering, N.} \emph{et~al.}
\newblock \bibinfo{journal}{\bibinfo{title}{Asymmetry in photoelectron emission
  from chiral molecules induced by circularly polarized light}}.
\newblock {\emph{\JournalTitle{Phys. Rev. Lett.}}}
  \textbf{\bibinfo{volume}{86}}, \bibinfo{pages}{1187--1190},
  \doiprefix\url{10.1103/PhysRevLett.86.1187} (\bibinfo{year}{2001}).

\bibitem{Comby2016}
\bibinfo{author}{Comby, A.} \emph{et~al.}
\newblock \bibinfo{journal}{\bibinfo{title}{Relaxation dynamics in photoexcited
  chiral molecules studied by time-resolved photoelectron circular dichroism:
  Toward chiral femtochemistry}}.
\newblock {\emph{\JournalTitle{Journal of Physical Chemistry Letters}}}
  \textbf{\bibinfo{volume}{7}}, \bibinfo{pages}{4514--4519},
  \doiprefix\url{10.1021/acs.jpclett.6b02065} (\bibinfo{year}{2016}).

\bibitem{Beauvarlet2022}
\bibinfo{author}{Beauvarlet, S.} \emph{et~al.}
\newblock \bibinfo{journal}{\bibinfo{title}{Photoelectron elliptical dichroism
  spectroscopy of resonance-enhanced multiphoton ionization via the 3s{,} 3p
  and 3d {R}ydberg series in fenchone}}.
\newblock {\emph{\JournalTitle{Phys. Chem. Chem. Phys.}}}
  \textbf{\bibinfo{volume}{24}}, \bibinfo{pages}{6415--6427},
  \doiprefix\url{10.1039/D1CP05618B} (\bibinfo{year}{2022}).

\bibitem{Baykusheva2019}
\bibinfo{author}{Baykusheva, D.} \emph{et~al.}
\newblock \bibinfo{journal}{\bibinfo{title}{Real-time probing of chirality
  during a chemical reaction}}.
\newblock {\emph{\JournalTitle{Proceedings of the National Academy of
  Sciences}}} \textbf{\bibinfo{volume}{116}}, \bibinfo{pages}{23923--23929},
  \doiprefix\url{10.1073/pnas.1907189116} (\bibinfo{year}{2019}).

\bibitem{Ilchen2021}
\bibinfo{author}{Ilchen, M.} \emph{et~al.}
\newblock \bibinfo{journal}{\bibinfo{title}{Site-specific interrogation of an
  ionic chiral fragment during photolysis using an {X}-ray free-electron
  laser}}.
\newblock {\emph{\JournalTitle{Communications Chemistry}}}
  \textbf{\bibinfo{volume}{4}}, \bibinfo{pages}{119},
  \doiprefix\url{10.1038/s42004-021-00555-6} (\bibinfo{year}{2021}).

\bibitem{Nugentglandorf2001}
\bibinfo{author}{Nugent-Glandorf, L.} \emph{et~al.}
\newblock \bibinfo{journal}{\bibinfo{title}{Ultrafast time-resolved soft
  {X}-ray photoelectron spectroscopy of dissociating {B}r$_2$}}.
\newblock {\emph{\JournalTitle{Phys. Rev. Lett.}}}
  \textbf{\bibinfo{volume}{87}}, \bibinfo{pages}{193002},
  \doiprefix\url{10.1103/PhysRevLett.87.193002} (\bibinfo{year}{2001}).

\bibitem{Squibb2018}
\bibinfo{author}{Squibb, R.~J.} \emph{et~al.}
\newblock \bibinfo{journal}{\bibinfo{title}{Acetylacetone photodynamics at a
  seeded free-electron laser}}.
\newblock {\emph{\JournalTitle{Nature Communications}}}
  \textbf{\bibinfo{volume}{9}}, \bibinfo{pages}{63} (\bibinfo{year}{2018}).

\bibitem{Conta2018}
\bibinfo{author}{von Conta, A.} \emph{et~al.}
\newblock \bibinfo{journal}{\bibinfo{title}{Conical-intersection dynamics and
  ground-state chemistry probed by extreme-ultraviolet time-resolved
  photoelectron spectroscopy}}.
\newblock {\emph{\JournalTitle{Nature Communications}}}
  \textbf{\bibinfo{volume}{9}}, \bibinfo{pages}{1--10},
  \doiprefix\url{10.1038/s41467-018-05292-4} (\bibinfo{year}{2018}).

\bibitem{Svoboda2019}
\bibinfo{author}{Svoboda, V.}, \bibinfo{author}{Wang, C.},
  \bibinfo{author}{Waters, M. D.~J.} \& \bibinfo{author}{W\"{o}rner, H.~J.}
\newblock \bibinfo{journal}{\bibinfo{title}{Electronic and vibrational
  relaxation dynamics of \ce{NH3} {R}ydberg states probed by vacuum-ultraviolet
  time-resolved photoelectron imaging}}.
\newblock {\emph{\JournalTitle{Journal of Chemical Physics}}}
  \textbf{\bibinfo{volume}{151}}, \bibinfo{pages}{104306},
  \doiprefix\url{10.1063/1.5116707} (\bibinfo{year}{2019}).

\bibitem{Powis2008}
\bibinfo{author}{Powis, I.}
\newblock \bibinfo{journal}{\bibinfo{title}{Photoelectron circular dichroism:
  Chiral asymmetry in the angular distribution of electrons emitted by
  (+)-\textit{S}-carvone}}.
\newblock {\emph{\JournalTitle{Chirality}}} \textbf{\bibinfo{volume}{20}},
  \bibinfo{pages}{961--968}, \doiprefix\url{10.1002/chir.20537}
  (\bibinfo{year}{2008}).

\bibitem{Powis2014}
\bibinfo{author}{Powis, I.}
\newblock \bibinfo{journal}{\bibinfo{title}{Communication: The influence of
  vibrational parity in chiral photoionization dynamics}}.
\newblock {\emph{\JournalTitle{Journal of Chemical Physics}}}
  \textbf{\bibinfo{volume}{140}}, \bibinfo{pages}{111103},
  \doiprefix\url{10.1063/1.4869204} (\bibinfo{year}{2014}).

\bibitem{Powis2014a}
\bibinfo{author}{Powis, I.} \emph{et~al.}
\newblock \bibinfo{journal}{\bibinfo{title}{A photoionization investigation of
  small, homochiral clusters of glycidol using circularly polarized radiation
  and velocity map electron{\textendash}ion coincidence imaging}}.
\newblock {\emph{\JournalTitle{Phys. Chem. Chem. Phys.}}}
  \textbf{\bibinfo{volume}{16}}, \bibinfo{pages}{467--476},
  \doiprefix\url{10.1039/c3cp53248h} (\bibinfo{year}{2014}).

\bibitem{Powis2008a}
\bibinfo{author}{Powis, I.} \emph{et~al.}
\newblock \bibinfo{journal}{\bibinfo{title}{Chiral asymmetry in the
  angle-resolved {O} and {C} 1s$^{-1}$ core photoemissions of the {R}
  enantiomer of glycidol}}.
\newblock {\emph{\JournalTitle{Physical Review A}}}
  \textbf{\bibinfo{volume}{78}}, \doiprefix\url{10.1103/physreva.78.052501}
  (\bibinfo{year}{2008}).

\bibitem{Garcia2014}
\bibinfo{author}{Garcia, G.~A.}, \bibinfo{author}{Dossmann, H.},
  \bibinfo{author}{Nahon, L.}, \bibinfo{author}{Daly, S.} \&
  \bibinfo{author}{Powis, I.}
\newblock \bibinfo{journal}{\bibinfo{title}{Photoelectron circular dichroism
  and spectroscopy of trifluoromethyl- and methyl-oxirane: a comparative
  study}}.
\newblock {\emph{\JournalTitle{Physical Chemistry Chemical Physics}}}
  \textbf{\bibinfo{volume}{16}}, \bibinfo{pages}{16214},
  \doiprefix\url{10.1039/c4cp01941e} (\bibinfo{year}{2014}).

\bibitem{Garcia2013}
\bibinfo{author}{Garcia, G.~A.}, \bibinfo{author}{Nahon, L.},
  \bibinfo{author}{Daly, S.} \& \bibinfo{author}{Powis, I.}
\newblock \bibinfo{journal}{\bibinfo{title}{Vibrationally induced inversion of
  photoelectron forward-backward asymmetry in chiral molecule photoionization
  by circularly polarized light}}.
\newblock {\emph{\JournalTitle{Nature Communications}}}
  \textbf{\bibinfo{volume}{4}}, \bibinfo{pages}{1--6},
  \doiprefix\url{10.1038/ncomms3132} (\bibinfo{year}{2013}).

\bibitem{Hergenhahn2004}
\bibinfo{author}{Hergenhahn, U.} \emph{et~al.}
\newblock \bibinfo{journal}{\bibinfo{title}{Photoelectron circular dichroism in
  core level ionization of randomly oriented pure enantiomers of the chiral
  molecule camphor}}.
\newblock {\emph{\JournalTitle{Journal of Chemical Physics}}}
  \textbf{\bibinfo{volume}{120}}, \bibinfo{pages}{4553--4556},
  \doiprefix\url{10.1063/1.1651474} (\bibinfo{year}{2004}).

\bibitem{Lux2016}
\bibinfo{author}{Lux, C.}, \bibinfo{author}{Senftleben, A.},
  \bibinfo{author}{Sarpe, C.}, \bibinfo{author}{Wollenhaupt, M.} \&
  \bibinfo{author}{Baumert, T.}
\newblock \bibinfo{journal}{\bibinfo{title}{Photoelectron circular dichroism
  observed in the above-threshold ionization signal from chiral molecules with
  femtosecond laser pulses}}.
\newblock {\emph{\JournalTitle{Journal of Physics B: Atomic, Molecular and
  Optical Physics}}} \textbf{\bibinfo{volume}{49}}, \bibinfo{pages}{02LT01},
  \doiprefix\url{10.1088/0953-4075/49/2/02lt01} (\bibinfo{year}{2015}).

\bibitem{Lux2015}
\bibinfo{author}{Lux, C.}, \bibinfo{author}{Wollenhaupt, M.},
  \bibinfo{author}{Sarpe, C.} \& \bibinfo{author}{Baumert, T.}
\newblock \bibinfo{journal}{\bibinfo{title}{Photoelectron circular dichroism of
  bicyclic ketones from multiphoton ionization with femtosecond laser pulses}}.
\newblock {\emph{\JournalTitle{{ChemPhysChem}}}} \textbf{\bibinfo{volume}{16}},
  \bibinfo{pages}{115--137}, \doiprefix\url{10.1002/cphc.201402643}
  (\bibinfo{year}{2014}).

\bibitem{Lux2012}
\bibinfo{author}{Lux, C.} \emph{et~al.}
\newblock \bibinfo{journal}{\bibinfo{title}{Circular dichroism in the
  photoelectron angular distributions of camphor and fenchone from multiphoton
  ionization with femtosecond laser pulses}}.
\newblock {\emph{\JournalTitle{Angewandte Chemie International Edition}}}
  \textbf{\bibinfo{volume}{51}}, \bibinfo{pages}{5001--5005},
  \doiprefix\url{10.1002/anie.201109035} (\bibinfo{year}{2012}).

\bibitem{Miles2017}
\bibinfo{author}{Miles, J.} \emph{et~al.}
\newblock \bibinfo{journal}{\bibinfo{title}{A new technique for probing
  chirality via photoelectron circular dichroism}}.
\newblock {\emph{\JournalTitle{Analytica Chimica Acta}}}
  \textbf{\bibinfo{volume}{984}}, \bibinfo{pages}{134--139},
  \doiprefix\url{10.1016/j.aca.2017.06.051} (\bibinfo{year}{2017}).

\bibitem{Kastner2016}
\bibinfo{author}{Kastner, A.} \emph{et~al.}
\newblock \bibinfo{journal}{\bibinfo{title}{Enantiomeric excess sensitivity to
  below one percent by using femtosecond photoelectron circular dichroism}}.
\newblock {\emph{\JournalTitle{{ChemPhysChem}}}} \textbf{\bibinfo{volume}{17}},
  \bibinfo{pages}{1119--1122}, \doiprefix\url{10.1002/cphc.201501067}
  (\bibinfo{year}{2016}).

\bibitem{Kastner2017}
\bibinfo{author}{Kastner, A.} \emph{et~al.}
\newblock \bibinfo{journal}{\bibinfo{title}{Intermediate state dependence of
  the photoelectron circular dichroism of fenchone observed via femtosecond
  resonance-enhanced multi-photon ionization}}.
\newblock {\emph{\JournalTitle{Journal of Chemical Physics}}}
  \textbf{\bibinfo{volume}{147}}, \bibinfo{pages}{013926},
  \doiprefix\url{10.1063/1.4982614} (\bibinfo{year}{2017}).

\bibitem{Goetz2017}
\bibinfo{author}{Goetz, R.~E.}, \bibinfo{author}{Isaev, T.~A.},
  \bibinfo{author}{Nikoobakht, B.}, \bibinfo{author}{Berger, R.} \&
  \bibinfo{author}{Koch, C.~P.}
\newblock \bibinfo{journal}{\bibinfo{title}{Theoretical description of circular
  dichroism in photoelectron angular distributions of randomly oriented chiral
  molecules after multi-photon photoionization}}.
\newblock {\emph{\JournalTitle{The Journal of Chemical Physics}}}
  \textbf{\bibinfo{volume}{146}}, \bibinfo{pages}{024306},
  \doiprefix\url{10.1063/1.4973456} (\bibinfo{year}{2017}).
\newblock \eprint{https://doi.org/10.1063/1.4973456}.

\bibitem{Artemyev2015}
\bibinfo{author}{Artemyev, A.~N.}, \bibinfo{author}{Müller, A.~D.},
  \bibinfo{author}{Hochstuhl, D.} \& \bibinfo{author}{Demekhin, P.~V.}
\newblock \bibinfo{journal}{\bibinfo{title}{Photoelectron circular dichroism in
  the multiphoton ionization by short laser pulses. {I}. propagation of
  single-active-electron wave packets in chiral pseudo-potentials}}.
\newblock {\emph{\JournalTitle{The Journal of Chemical Physics}}}
  \textbf{\bibinfo{volume}{142}}, \bibinfo{pages}{244105},
  \doiprefix\url{10.1063/1.4922690} (\bibinfo{year}{2015}).
\newblock \eprint{https://doi.org/10.1063/1.4922690}.

\bibitem{Ulrich2008}
\bibinfo{author}{Ulrich, V.} \emph{et~al.}
\newblock \bibinfo{journal}{\bibinfo{title}{Giant chiral asymmetry in the {C}
  1s core level photoemission from randomly oriented fenchone enantiomers}}.
\newblock {\emph{\JournalTitle{Journal of Physical Chemistry A}}}
  \textbf{\bibinfo{volume}{112}}, \bibinfo{pages}{3544--3549},
  \doiprefix\url{10.1021/jp709761u} (\bibinfo{year}{2008}).

\bibitem{Lehmann2013}
\bibinfo{author}{Lehmann, C.~S.}, \bibinfo{author}{Ram, N.~B.},
  \bibinfo{author}{Powis, I.} \& \bibinfo{author}{Janssen, M. H.~M.}
\newblock \bibinfo{journal}{\bibinfo{title}{Imaging photoelectron circular
  dichroism of chiral molecules by femtosecond multiphoton coincidence
  detection}}.
\newblock {\emph{\JournalTitle{Journal of Chemical Physics}}}
  \textbf{\bibinfo{volume}{139}}, \bibinfo{pages}{234307},
  \doiprefix\url{10.1063/1.4844295} (\bibinfo{year}{2013}).

\bibitem{Nahon2016}
\bibinfo{author}{Nahon, L.} \emph{et~al.}
\newblock \bibinfo{journal}{\bibinfo{title}{Determination of accurate electron
  chiral asymmetries in fenchone and camphor in the {VUV} range: sensitivity to
  isomerism and enantiomeric purity}}.
\newblock {\emph{\JournalTitle{Physical Chemistry Chemical Physics}}}
  \textbf{\bibinfo{volume}{18}}, \bibinfo{pages}{12696--12706},
  \doiprefix\url{10.1039/c6cp01293k} (\bibinfo{year}{2016}).

\bibitem{Comby2018}
\bibinfo{author}{Comby, A.} \emph{et~al.}
\newblock \bibinfo{journal}{\bibinfo{title}{Real-time determination of
  enantiomeric and isomeric content using photoelectron elliptical dichroism}}.
\newblock {\emph{\JournalTitle{Nature Communications}}}
  \textbf{\bibinfo{volume}{9}}, \bibinfo{pages}{5212},
  \doiprefix\url{10.1038/s41467-018-07609-9} (\bibinfo{year}{2018}).

\bibitem{Blanchet2021}
\bibinfo{author}{Blanchet, V.} \emph{et~al.}
\newblock \bibinfo{journal}{\bibinfo{title}{Ultrafast relaxation investigated
  by photoelectron circular dichroism: an isomeric comparison of camphor and
  fenchone}}.
\newblock {\emph{\JournalTitle{Phys. Chem. Chem. Phys.}}}
  \textbf{\bibinfo{volume}{23}}, \bibinfo{pages}{25612--25628},
  \doiprefix\url{10.1039/D1CP03569J} (\bibinfo{year}{2021}).

\bibitem{Powis2008c}
\bibinfo{author}{Powis, I.}, \bibinfo{author}{Harding, C.~J.},
  \bibinfo{author}{Garcia, G.~A.} \& \bibinfo{author}{Nahon, L.}
\newblock \bibinfo{journal}{\bibinfo{title}{A valence photoelectron imaging
  investigation of chiral asymmetry in the photoionization of fenchone and
  camphor}}.
\newblock {\emph{\JournalTitle{ChemPhysChem}}} \textbf{\bibinfo{volume}{9}},
  \bibinfo{pages}{475--483} (\bibinfo{year}{2008}).

\bibitem{Stener2004}
\bibinfo{author}{Stener, M.}, \bibinfo{author}{Fronzoni, G.},
  \bibinfo{author}{Tommaso, D.~D.} \& \bibinfo{author}{Decleva, P.}
\newblock \bibinfo{journal}{\bibinfo{title}{Density functional study on the
  circular dichroism of photoelectron angular distribution from chiral
  derivatives of oxirane}}.
\newblock {\emph{\JournalTitle{Journal of Chemical Physics}}}
  \textbf{\bibinfo{volume}{120}}, \bibinfo{pages}{3284--3296},
  \doiprefix\url{10.1063/1.1640617} (\bibinfo{year}{2004}).

\bibitem{Natalense1999}
\bibinfo{author}{Natalense, A. P.~P.} \& \bibinfo{author}{Lucchese, R.~R.}
\newblock \bibinfo{journal}{\bibinfo{title}{Cross section and asymmetry
  parameter calculation for sulfur 1s photoionization of {SF}$_6$}}.
\newblock {\emph{\JournalTitle{Journal of Chemical Physics}}}
  \textbf{\bibinfo{volume}{111}}, \bibinfo{pages}{5344--5348},
  \doiprefix\url{10.1063/1.479794} (\bibinfo{year}{1999}).

\bibitem{Gianturco1994}
\bibinfo{author}{Gianturco, F.~A.}, \bibinfo{author}{Lucchese, R.~R.} \&
  \bibinfo{author}{Sanna, N.}
\newblock \bibinfo{journal}{\bibinfo{title}{Calculation of low-energy elastic
  cross sections for electron-{CF}$_4$ scattering}}.
\newblock {\emph{\JournalTitle{Journal of Chemical Physics}}}
  \textbf{\bibinfo{volume}{100}}, \bibinfo{pages}{6464--6471},
  \doiprefix\url{10.1063/1.467237} (\bibinfo{year}{1994}).

\bibitem{Svoboda2022}
\bibinfo{author}{Svoboda, V.}, \bibinfo{author}{Waters, M. D.~J.},
  \bibinfo{author}{Zindel, D.} \& \bibinfo{author}{W\"{o}rner, H.~J.}
\newblock \bibinfo{journal}{\bibinfo{title}{Generation and complete polarimetry
  of ultrashort circularly polarized extreme-ultraviolet pulses}}.
\newblock {\emph{\JournalTitle{Optics Express}}} \textbf{\bibinfo{volume}{30}},
  \bibinfo{pages}{14358--14367}, \doiprefix\url{10.1364/OE.449411}
  (\bibinfo{year}{2022}).

\bibitem{Novak2002}
\bibinfo{author}{Novak, I.}, \bibinfo{author}{Li, D.~B.} \&
  \bibinfo{author}{Potts, A.~W.}
\newblock \bibinfo{journal}{\bibinfo{title}{Electronic structure of chiral
  halomethanes}}.
\newblock {\emph{\JournalTitle{Journal of Physical Chemistry A}}}
  \textbf{\bibinfo{volume}{106}}, \bibinfo{pages}{465--468},
  \doiprefix\url{10.1021/jp0116959} (\bibinfo{year}{2002}).

\bibitem{He2009}
\bibinfo{author}{He, Y.-L.} \& \bibinfo{author}{Wang, L.}
\newblock \bibinfo{journal}{\bibinfo{title}{Cations of halogenated methanes:
  adiabatic ionization energies, potential energy surfaces, and ion fragment
  appearance energies}}.
\newblock {\emph{\JournalTitle{Structural Chemistry}}}
  \textbf{\bibinfo{volume}{20}}, \bibinfo{pages}{461--479},
  \doiprefix\url{10.1007/s11224-009-9444-x} (\bibinfo{year}{2009}).

\bibitem{Imre1980}
\bibinfo{author}{Imre, D.} \& \bibinfo{author}{Koenig, T.}
\newblock \bibinfo{journal}{\bibinfo{title}{The {H}e({I}) photoelectron
  spectrum of atomic iodine by photodissociation of molecular iodine}}.
\newblock {\emph{\JournalTitle{Chemical Physics Letters}}}
  \textbf{\bibinfo{volume}{73}}, \bibinfo{pages}{62--66},
  \doiprefix\url{10.1016/0009-2614(80)85203-1} (\bibinfo{year}{1980}).

\bibitem{Powis2008b}
\bibinfo{author}{Powis, I.}
\newblock \bibinfo{title}{Photoelectron circular dichroism in chiral
  molecules}.
\newblock In \emph{\bibinfo{booktitle}{Advances in Chemical Physics}},
  \bibinfo{pages}{267--329}, \doiprefix\url{10.1002/9780470259474.ch5}
  (\bibinfo{publisher}{John Wiley {\&} Sons, Inc.}, \bibinfo{year}{2008}).

\bibitem{Schultz1984}
\bibinfo{author}{Schultz, J.~C.}, \bibinfo{author}{Houle, F.~A.} \&
  \bibinfo{author}{Beauchamp, J.~L.}
\newblock \bibinfo{journal}{\bibinfo{title}{Photoelectron spectroscopy of
  1-propyl, 1-butyl, isobutyl, neopentyl, and 2-butyl radicals: free radical
  precursors to high-energy carbonium ion isomers}}.
\newblock {\emph{\JournalTitle{Journal of the American Chemical Society}}}
  \textbf{\bibinfo{volume}{106}}, \bibinfo{pages}{3917--3927},
  \doiprefix\url{10.1021/ja00326a006} (\bibinfo{year}{1984}).

\bibitem{Park2001}
\bibinfo{author}{Park, S.~T.}, \bibinfo{author}{Kim, S.~K.} \&
  \bibinfo{author}{Kim, M.~S.}
\newblock \bibinfo{journal}{\bibinfo{title}{Vacuum-ultraviolet mass-analyzed
  threshold ionization spectra of iodobutane isomers: Conformer-specific
  ionization and ion-core dissociation followed by ionization}}.
\newblock {\emph{\JournalTitle{Journal of Chemical Physics}}}
  \textbf{\bibinfo{volume}{115}}, \bibinfo{pages}{2492--2498},
  \doiprefix\url{10.1063/1.1386786} (\bibinfo{year}{2001}).

\bibitem{Corrales2014}
\bibinfo{author}{Corrales, M.~E.} \emph{et~al.}
\newblock \bibinfo{journal}{\bibinfo{title}{Structural dynamics effects on the
  ultrafast chemical bond cleavage of a photodissociation reaction}}.
\newblock {\emph{\JournalTitle{Physical Chemistry Chemical Physics}}}
  \textbf{\bibinfo{volume}{16}}, \bibinfo{pages}{8812},
  \doiprefix\url{10.1039/c3cp54677b} (\bibinfo{year}{2014}).

\bibitem{Werner2011}
\bibinfo{author}{Werner, H.-J.}, \bibinfo{author}{Knowles, P.~J.},
  \bibinfo{author}{Knizia, G.}, \bibinfo{author}{Manby, F.~R.} \&
  \bibinfo{author}{Schütz, M.}
\newblock \bibinfo{journal}{\bibinfo{title}{Molpro: a general-purpose quantum
  chemistry program package}}.
\newblock {\emph{\JournalTitle{Wiley Interdisciplinary Reviews: Computational
  Molecular Science}}} \textbf{\bibinfo{volume}{2}}, \bibinfo{pages}{242--253},
  \doiprefix\url{10.1002/wcms.82} (\bibinfo{year}{2011}).

\bibitem{Molpro2015}
\bibinfo{author}{Werner, H.-J.} \emph{et~al.}
\newblock \bibinfo{title}{Molpro, version 2015.1, a package of \textit{ab
  initio} programs}.

\bibitem{Quack1986}
\bibinfo{author}{Quack, M.}
\newblock \bibinfo{journal}{\bibinfo{title}{On the measurement of the parity
  violating energy difference between enantiomers}}.
\newblock {\emph{\JournalTitle{Chemical Physics Letters}}}
  \textbf{\bibinfo{volume}{132}}, \bibinfo{pages}{147--153},
  \doiprefix\url{10.1016/0009-2614(86)80098-7} (\bibinfo{year}{1986}).

\bibitem{Colbert1992}
\bibinfo{author}{Colbert, D.~T.} \& \bibinfo{author}{Miller, W.~H.}
\newblock \bibinfo{journal}{\bibinfo{title}{A novel discrete variable
  representation for quantum mechanical reactive scattering via the {S}-matrix
  {K}ohn method}}.
\newblock {\emph{\JournalTitle{Journal of Chemical Physics}}}
  \textbf{\bibinfo{volume}{96}}, \bibinfo{pages}{1982--1991},
  \doiprefix\url{10.1063/1.462100} (\bibinfo{year}{1992}).

\bibitem{Svoboda2017}
\bibinfo{author}{Svoboda, V.}, \bibinfo{author}{Ram, N.~B.},
  \bibinfo{author}{Rajeev, R.} \& \bibinfo{author}{W{\"{o}}rner, H.~J.}
\newblock \bibinfo{journal}{\bibinfo{title}{Time-resolved photoelectron imaging
  with a femtosecond vacuum-ultraviolet light source: Dynamics in the
  $\tilde{A}$/$\tilde{B}$- and $\tilde{F}$-bands of {SO}$_2$}}.
\newblock {\emph{\JournalTitle{Journal of Chemical Physics}}}
  \textbf{\bibinfo{volume}{146}}, \bibinfo{pages}{084301},
  \doiprefix\url{10.1063/1.4976552} (\bibinfo{year}{2017}).

\end{thebibliography}

\end{document}